# New Diluted Ferromagnetic Semiconductor isostructural to "122" type iron pnictide superconductor with $T_C$ up to 180 K

K. Zhao[1], Z. Deng[1], X. C. Wang[1], W. Han[1], J. L. Zhu[1], X. Li[1], Q.Q. Liu[1], R.C. Yu[1], T. Goko[2], B. Frandsen[2], Lian Liu[2], Fanlong Ning[2,3], Y.J. Uemura[2], H. Dabkowska[4], G.M. Luke[4], H. Luetkens[5], E. Morenzoni[5], S.R. Dunsiger[6], A. Senyshyn[6], P. Böni[6], and C.Q. Jin[1]

1) *Beijing National Laboratory for Condensed Matter Physics, and Institute of Physics, Chinese Academy of Sciences, Beijing 100190, China*
2) *Department of Physics, Columbia University, New York, New York 10027, USA*
3) *Department of Physics, Zhejiang University, Hangzhou 310027, China*
4) *Department of Physics and Astronomy, McMaster University, Hamilton, Ontario L8S 4M1, Canada*
5) *Paul Scherrer Institute, Laboratory for Muon Spin Spectroscopy, CH-5232 Villigen PSI, Switzerland*
6) *Physics Department and FRM-II, Technische Universität München, D-85748 Garching. München, Germany*

**Diluted magnetic semiconductors (DMS) have received much attention due to its potential applications to spintronics devices. A prototypical system (Ga,Mn)As has been widely studied since 1990s. The simultaneous spin and charge doping via hetero-valence ($Ga^{3+}$,$Mn^{2+}$) substitution, however, resulted in severely limited solubility without availability of bulk specimens. Previously we synthesized a new diluted ferromagnetic semiconductor of bulk Li(Zn,Mn)As with Tc up to 50K, where isovalent (Zn,Mn) spin doping was separated from charge control via Li concentrations. Here we report the synthesis of a new diluted ferromagnetic semiconductor $(Ba_{1-x}K_x)(Zn_{1-y}Mn_y)_2As_2$, isostructural to iron 122 system, where holes are doped via ($Ba^{2+}$, $K^{1+}$), while spins via ($Zn^{2+}$,$Mn^{2+}$) substitutions. Bulk samples with x=0.1-0.3 and y=0.05-0.15 exhibit ferromagnetic order with $T_C$ up to 180K, comparable to that of record high Tc for Ga(MnAs), significantly enhanced than Li(Zn,Mn)As. Moreover the $(Ba,K)(Zn,Mn)_2As_2$ shares the same 122 crystal structure with semiconducting $BaZn_2As_2$, antiferromagnetic $BaMn_2As_2$, and superconducting $(Ba,K)Fe_2As_2$, which makes them promising to the development of multilayer functional devices.**

Diluted (or Doped) Magnetic Semiconductors (DMS) have received much attention due to their potential application in the field of spin-sensitive electronics (spintronics) [1-5]. DMS systems are produced by doping semiconductors with magnetic metal elements. In typical systems based on III-V semiconductors, such as (Ga,Mn)As, (In,Mn)As and (Ga,Mn)N, substitution of divalent Mn atoms into trivalent Ga (or In) sites leads to severely limited chemical solubility, resulting in metastable specimens only available as epitaxial thin films. The hetero-valence substitution, which simultaneously dopes both hole-carriers and magnetic atoms, makes it difficult to individually control charge and spin concentrations for better tuning of quantum freedom.

Following a theoretical proposal by Masek et al. [6], a new system Li(Zn,Mn)As was recently discovered by Deng et al. [7] based on the I-II-V semiconductor LiZnAs, showing a Curie temperature up to $T_C$ = 50 K. In this system, charges are doped via off-stoichiometry of Li concentrations, while spins by the isovalent ($Zn^{2+}$,$Mn^{2+}$) substitutions. Although Li(Zn,Mn)As was a ferromagnetic DMS of a new type having a few distinct advantages over (Ga,Mn)As, the upper limit of currently achievable $T_C$ has been significantly lower than that in (Ga,Mn)As [2,7].

$BaZn_2As_2$ [8] is a semiconductor with tetragonal $ThCr_2Si_2$ crystal structure synthetized at high temperature (>900°C) (shown in Fig. 1(a)), identical to that of $BaFe_2As_2$ [9], $BaMn_2As_2$ [10,11] and $(Ba,K)Mn_2As_2$ [12]. $(Ba,K)Fe_2As_2$ [9,13] is a classic member of the "122" type iron pnictide superconductors with transition temperature up to 38K, while $BaMn_2As_2$ is an antiferromagnet with $T_N$ ~ 625K [11,12]. It is noted that the stable phase at room temperature of $BaZn_2As_2$ crystallizes into a different orthorhombic structure with space group Pnma. (Supplementary Figure S1) However we found that 10% of K or Mn doping dramatically stabilizes the tetragonal $ThCr_2Si_2$ structure at room temperature or above 3.5K as revealed by low temperature neutron diffractions in this work. (Supplementary Figure S2) Here we report synthesis of a new ferromagnetic DMS $(Ba,K)(Zn,Mn)_2As_2$ system which share the same "122" structure with all these relevant systems. Via (Ba,K) substitution to dope hole carriers and (Zn,Mn) substitution to supply magnetic moments, the systems with 5-15 % Mn doping exhibit ferromagnetic order with $T_C$ up to 180K.

Figure 1(b) shows the X-ray diffraction results of $(Ba_{1-x}K_x)(Zn_{0.9}Mn_{0.1})_2As_2$ for $x$ = 0.05, 0.1, 0.15, 0.2, 0.25, and 0.3, respectively. The patterns with a 2θ range (10° to 80°) were collected, and the least-squares method was used to determine the lattice parameters of all polycrystalline samples, as shown in Fig. 1(c). Compared with the lattice parameters $a$=4.121Å and $c$=13.575 Å of $BaZn_2As_2$, the $a$-axis expanded and the $c$-axis shrank in $(Ba_{1-x}K_x)Zn_2As_2$. For compounds with Mn 10%, $(Ba_{1-x}K_x)(Zn_{0.9}Mn_{0.1})_2As_2$, the $c$-axis has similar tendency with Mn free $(Ba_{1-x}K_x)Zn_2As_2$, while the $a$-axis does not show obvious variation tendency as doping. These results indicate successful solid solutions of K and Mn.

Further structural studies have been made by neutron diffraction measurements from powder specimen of $(Ba_{1-x}K_x)(Zn_{0.9}Mn_{0.1})_2As_2$ with x ~ 0.3 at the FRM-II reactor in TU Munich. The powder diffraction pattern, shown in Fig. 1(d), fitted well to the structure found by X-rays, and contrasting neutron scattering length of Mn and Zn allowed us to confirm random substitution of Mn atoms in Zn sites. The residuals of the Rietveld analysis were about 4-5 %. There was no structural phase transition below T ~ 300 K, which is in contrast to the case of antiferromagnetic $(Ba,K)Fe_2As_2$ [9]. Due to small average moment size (average 0.1 Bohr magneton per Zn/Mn site) and limited neutron intensity, we could not detect conclusive signal from ferromagnetic spins below $T_C$.

Resistivity measurements shown in Fig. 1(e) indicate that $BaZn_2As_2$ is a semiconductor. Doping K atoms into Ba sites introduces hole carriers, leading to metallic behaviors of $(Ba,K)Zn_2As_2$. The resistivity curves of $(Ba_{1-x}K_x)(Zn_{0.9}Mn_{0.1})_2As_2$, for selected values of x up to 0.3, exhibit a small increase at low temperatures due presumably to spin scattering of carriers caused by Mn dopants. This variation of resistivity, similar to that of (Ga,Mn)N [14], is often observed in heavily-doped semiconductors. Strictly metallic behavior (with monotonic decrease of resistivity with decreasing temperatures) is not a precondition of having a ferromagnetic coupling between Mn moments mediated by RKKY interaction, as discussed in [1] and [15].

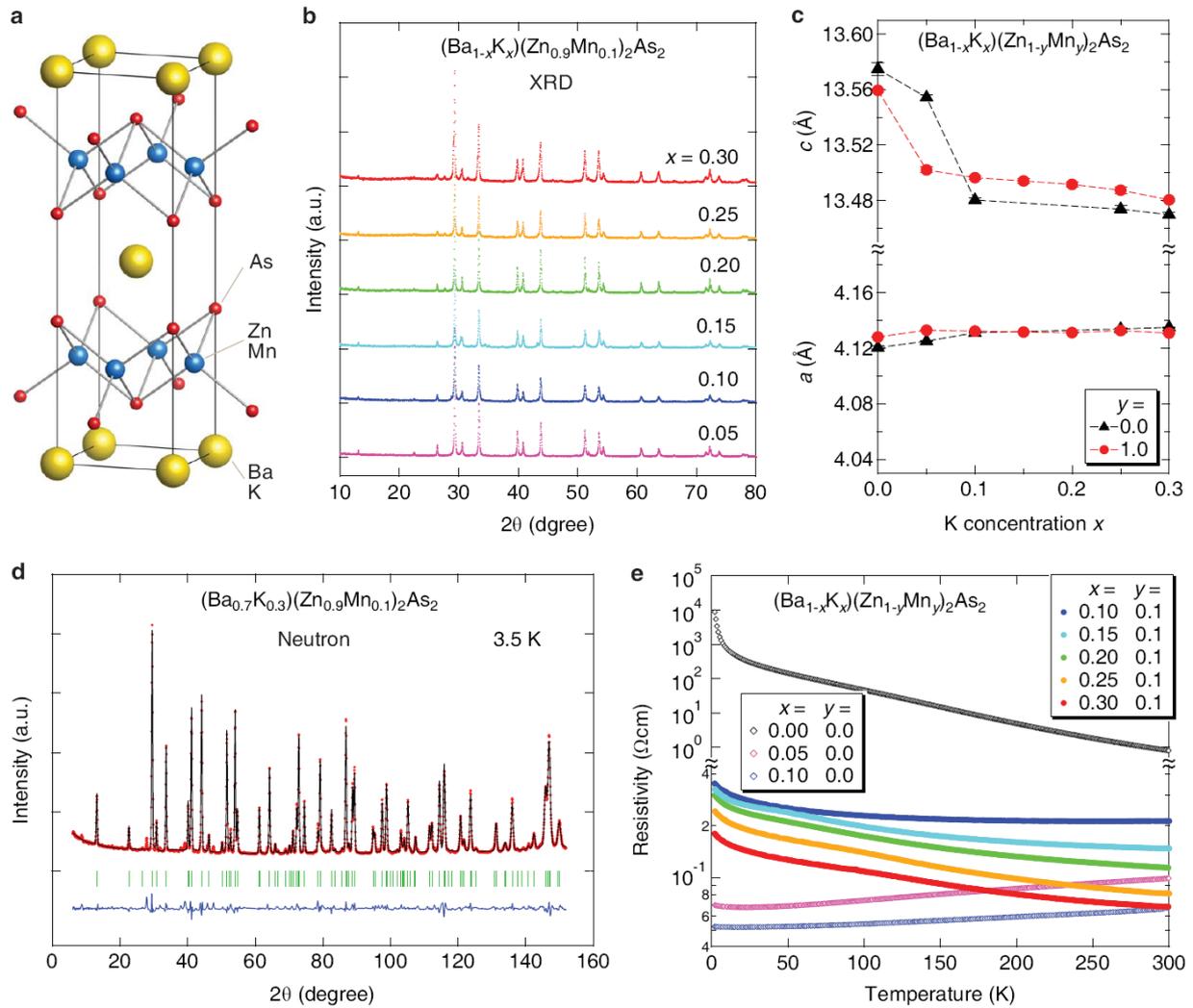

*Figure 1. (a) Crystal structure of (Ba,K)(Zn,Mn)$_2$As$_2$ belonging to tetragonal ThCr$_2$Si$_2$ structure. (b) X-ray diffraction (XRD) pattern of (Ba$_{1-x}$K$_x$)(Zn$_{0.9}$Mn$_{0.1}$)$_2$As$_2$ with several K concentrations x taken at room temperature. (c) c-axis and a-axis lattice constants obtained from XRD. (d) Neutron diffraction pattern from powder specimen of (Ba$_{1-x}$K$_x$)(Zn$_{0.9}$Mn$_{0.1}$)$_2$As$_2$ with x = 0.3 shown with Rietveld analyses. (e) Resistivity of (Ba$_{1-x}$K$_x$)(Zn$_{1-y}$Mn$_y$)$_2$As$_2$ for the pure Zn (y =0) and Mn 10% (y = 0.1) systems with several different charge doping levels x. The vertical axis for the pure BaZn$_2$As$_2$ is very different from that for other specimens.*

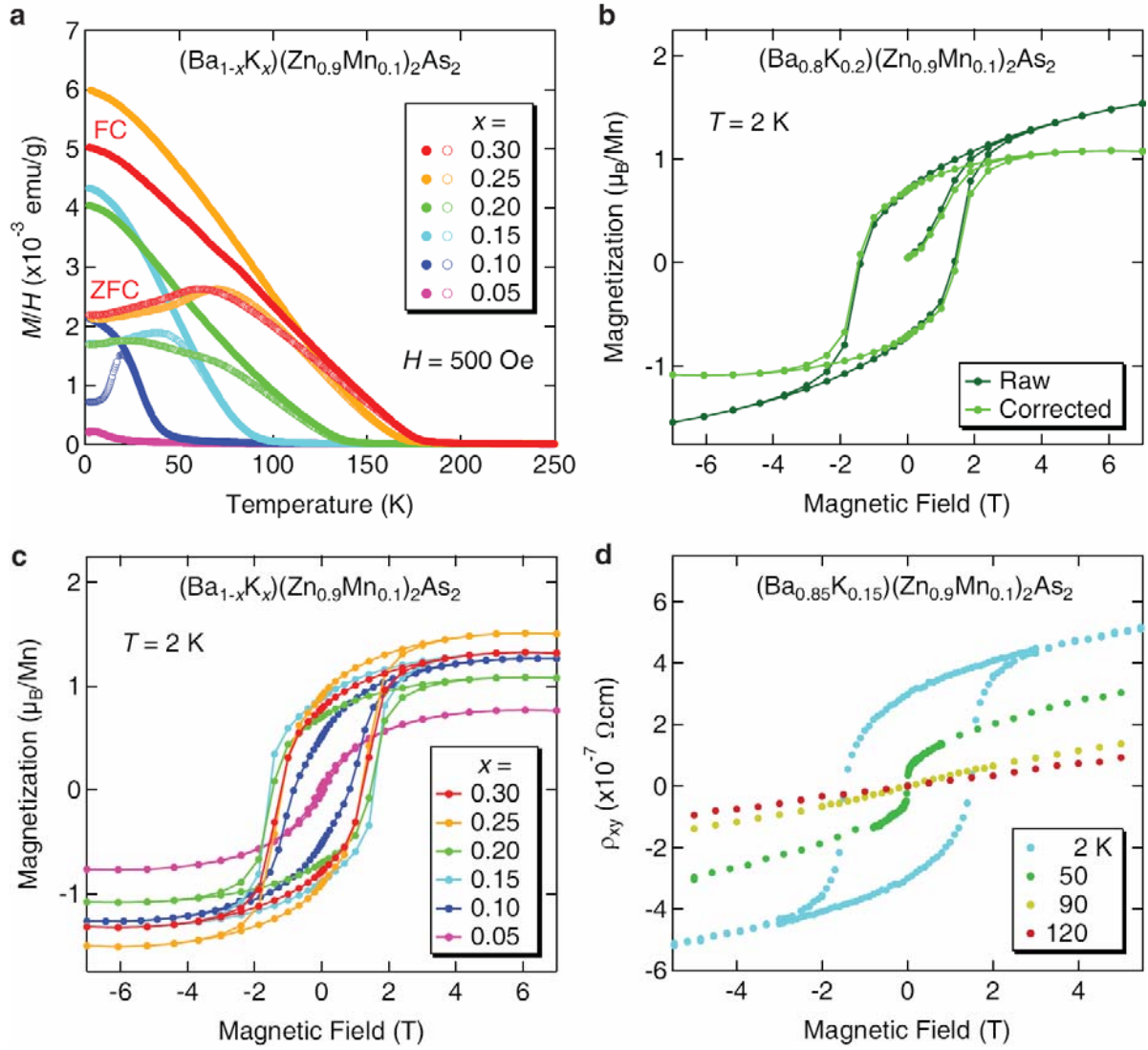

*Figure 2: (a) dc magnetization measured in H = 500 G in $(Ba_{1-x}K_x)(Zn_{0.9}Mn_{0.1})_2As_2$ with several different charge doping levels x, with zero-field cooling (ZFC) and field-cooling (FC) procedures. (b) Magnetic hysteresis curve M(H), measured in the external field H up to 7 T, showing initial magnetization curve and subtraction of a small T-linear contribution. (c) M(H) curves after field training and subtraction of the paramagnetic component measured in $(Ba_{1-x}K_x)(Zn_{0.9}Mn_{0.1})_2As$ with several different charge doping levels x. (d) Hall effect results from a sintered specimen of $(Ba_{1-x}K_x)(Zn_{0.9}Mn_{0.1})_2As_2$ with the charge doping level of x = 0.15 having $T_C$ ~ 90 K. Anomalous Hall effect and a very small coercive field is seen at T = 50 K near the history dependence temperature $T_{Hist}$, while a large coercive field is seen at T = 2 K.*

Figure 2(a) shows temperature dependence of magnetization in zero-field-cooling (ZFC) and field cooling (FC) procedures under 500Oe for $(Ba_{1-x}K_x)(Zn_{0.9}Mn_{0.1})_2As_2$ specimens with x = 0.05, 0.1, 0.15, 0.2, 0.25, and 0.3, respectively. Clear signatures of ferromagnetic order are seen in the curves, with corresponding critical temperature $T_C$ = 5K, 40K, 90K, 135K, 170K, and 180K, respectively. Above $T_C$, the samples are paramagnetic and the $\chi(T)$ curve can be fitted with the Curie-Weiss formula with the effective paramagnetic moment ~ $5\mu_B$ per $Mn^{2+}$. The hysteresis curves M(H) at T=2 K in Fig. 2(b) exhibit initial increase from H=0 state achieved by

ZFC procedure, and a small H-linear component, which is presumably due to remaining paramagnetic spins and/or field-induced polarization. By subtracting this small T-linear component, we obtain the M(H) curves of $(Ba_{1-x}K_x)(Zn_{0.9}Mn_{0.1})_2As_2$ at T = 2 K shown in Fig. 2(c). The saturation moment of 1 ~ 2 $\mu_B$ per Mn atom is comparable with that of (Ga,Mn)As [1] and Li(Zn,Mn)As [7].

In $(Ba,K)Zn_2As_2$, specimens with 10% (Ba,K) substitution results in hole concentration of $4.3*10^{20}$ cm$^{-3}$, consistent within a factor of two with that obtained by assuming that each K atom introduces one hole to the system. In $(Ba,K)(Zn,Mn)_2As_2$, linear dependence of Hall resistivity with magnetic field is observed above $T_C$. As shown in Fig. 2(d) for the x = 0.15 and y = 0.10 system, having $T_C$ =90 K, the Hall resistivity deviates from the linear dependence in low field at $T_C$. In the ferromagnetic state below $T_C$, the anomalous Hall effect is observed with a small coercive field ~ 35 Oe in the temperature region between $T_C$ and the "history dependence temperature" $T_{Hist}$ below which FC and ZFC susceptibility shows deviation. The small coercive field above $T_{Hist}$ will be helpful for spin manipulation.

The deviation between the magnetization in FC and ZFC procedures in Fig. 2(a) and a large difference between the initial and the field-trained M(H) curves in Fig. 2(b) can be ascribed to a large coercive force demonstrated in Fig. 2(c) below $T_{Hist}$. A very similar departure of ZFC and FC magnetization, associated with a coercive field ~ 1 T, was observed in a well-known itinerant ferromagnet SrRuO$_3$ [16,17], which has a comparable $T_C$ ~ 160K. The large coercive field at low temperatures in $(Ba,K)(Zn,Mn)_2As_2$ is different from Li(Zn,Mn)As which does not exhibit large coercive field in any temperature range. The large temperature dependence of the coercive field in $(Ba,K)(Zn,Mn)_2As_2$ opens a possibility of temperature tuning of anisotropy and stability of magnetic memory.

Using bulk polycrystalline specimens, we also performed positive muon spin relaxation measurements at Paul Scherrer Institute. Figure 3(a) shows the time spectra of the zero-field (ZF) MuSR on a $(Ba_{0.8}K_{0.2})(Zn_{0.9}Mn_{0.1})_2As_2$ specimen which has $T_C$ ~ 140 K, as determined by magnetization (Inset of Figure 3(b)). A sharp increase of the muon spin relaxation rate is seen with decreasing temperature below $T_C$, and a coherent precession signal was observed below T ~ 40 K. The volume fraction of magnetically ordered state was estimated by using MuSR data in ZF and weak transverse field (WTF) of 50 G as shown in Fig. 3(b). The MuSR results indicate static magnetic order developing in the entire volume with a rather sharp onset below $T_C$.

The ZF precession spectra in Fig. 3(a) at T = 5 and 20 K look very similar to that observed in the ferromagnetic SrRuO$_3$ by ZF-MuSR (see Fig. 7 of [18] and [19]). Small oscillation amplitudes in both systems may be due to domain structures and spread of demagnetizing field in polycrystalline specimens. We also note that ZF-MuSR spectra did not show oscillation in ferromagnetic Li(Zn,Mn)As [7] and (Ga,Mn)As [15]. In all these DMS systems, random substitution of Mn at Zn or Ga sites generates spatially random distribution of Mn moments, which makes the local field at the muon site highly random even in the ferromagnetic ground state. Due to this feature, ZF precession signals are strongly damped.

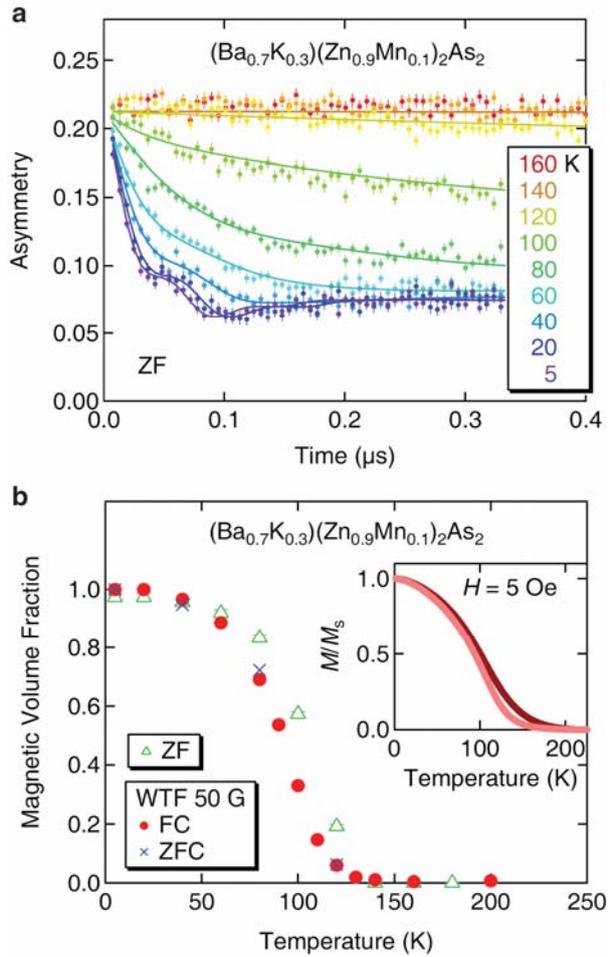

*Figure 3: (a) Zero-field MuSR time spectra obtained in polycrystalline specimen of $(Ba_{0.8}K_{0.2})(Zn_{0.9}Mn_{0.1})_2As_2$. (b) Volume fraction of regions with static magnetic order, estimated by MuSR measurements in zero field (ZF) and weak transverse field (WTF) of 50 G. No hysteresis is seen for measurements with ZFC and FC in 500 G. Inset: dc magnetization results of the specimens used in MuSR measurements.*

Table 1: Transition temperature $T_C$ and the saturation moment $M_S(H=0)$ at $T = 2$ K after training in the external field of 7 T for $Ba_{1-x}K_x(Zn_{1-y}Mn_y)_2As_2$ with several different Mn spin doping levels y and hole charge doping levels x.

| $Ba_{1-x}K_x(Zn_{1-y}Mn_y)_2As_2$ $T_C$[K] ( Saturation Moment $M_S$ [Bohr Magneton / Mn] ) | | | | | | |
|---|---|---|---|---|---|---|
| | x=0.05 | x=0.1 | x=0.15 | x=0.2 | x=0.25 | x=0.3 |
| y=0.05 | | 30(1.4) | 75(1.8) | 105(1.8) | 160(2.0) | 160(1.7) |
| y=0.1 | 5(0.8) | 40(1.3) | 90(1.3) | 135(1.1) | 170(1.5) | 180(1.3) |
| y=0.15 | 10(0.4) | 40(0.6) | 90(1.4) | 145(1.2) | 180(1.5) | 190(1.1) |

Table 2: Selected properties of (Ga,Mn)As, Li(Zn,Mn)As and (Ba,K)(Zn,Mn)$_2$As$_2$.

| | (Ga,Mn)As [21] | Li(Zn,Mn)As | (Ba,K)(Zn,Mn)$_2$As$_2$ |
|---|---|---|---|
| Valence before doping | III– V | I – II – V | II – II $_2$– V $_2$ |
| FM temperature | 185K | 50K | 180K |
| Coercieve field | <100Oe (5K) | 30-100Oe (2K) | 10000Oe (2K) |
| Saturation moment( /Mn) | 5$\mu_B$ | 2.9$\mu_B$ | 2$\mu_B$ |
| Sample form | epitaxial thin film | bulk specimen | bulk specimen |

In Table 1, we summarize transition temperature $T_C$ determined by ZFC and FC magnetization, and the size of the ordered moment per Mn at T = 2 K, obtained from the H= 0 values of the M(H) curve after cycling the field to 7 T, for $(Ba_{1-x}K_x)(Zn_{1-y}Mn_y)_2As_2$ with the K concentration x up to 0.3 and Mn concentration y up to 0.15. The highest $T_C$ is obtained for x = 0.3 and y = 0.1. We notice a tendency for the reduction of the moment size with increasing Mn doping, which may be a result from competition between antiferromagnetic coupling of Mn moment in the nearest neighbor Zn sites and ferromagnetic coupling between Mn moments in more distant locations mediated by the doped hole carriers. This feature is common to (Ga,Mn)As and Li(Zn,Mn)As.

The present system is different from Li(Zn,Mn)As in several aspects: (1) currently available highest value of $T_C$ is more than 3 times higher in (Ba,K)(Zn,Mn)$_2$As$_2$ than that in Li(Zn,Mn)As (2) the present system shows a very high coercive field at low temperatures. In addition, notable difference lies in (3) crystal structures. Crystal structure of the ``111'' DMS system Li(Zn,Mn)As, is different from that of the relevant antiferromagnet LiMnAs and also from superconducting LiFeAs, although they share a common square-lattice As layers. In the present ``122'' DMS ferromagnet (Ba,K)(Zn,Mn)$_2$As$_2$, semiconducting BaZn$_2$As$_2$, antiferromagnetic BaMn$_2$As$_2$ and superconducting (Ba,K)Fe$_2$As$_2$ all share the same crystal structure shown in Fig. 1(a), except for a small orthorhombic distortion in (Ba,K)Fe$_2$As$_2$ associated with antiferromagnetic order. Moreover, the lattice constants in the a-b plane matches within about 5% as shown in Table 2. These features could provide distinct advantages to the present system over the 111 DMS systems in attempts to generate functional devices based on junctions of various combinations of these states.

Table 3: Crystal structures and lattice constants of superconducting (Ba,K)Fe$_2$As$_2$, antiferromagnetic BaMn$_2$As$_2$, semiconducting BaZn$_2$As$_2$, and ferromagnetic (Ba,K)(Zn,Mn)$_2$As$_2$.

Possible research use of such junctions includes, for example, quantitative estimation of the carrier spin polarization and scattering strength via Andreev reflection at the DMS-

| Compound [ref] | (Ba,K)Fe$_2$As$_2$ [13] | BaMn$_2$As$_2$ [9,11] | BaZn$_2$As$_2$ | (Ba,K)(Zn,Mn)$_2$As$_2$ present work |
|---|---|---|---|---|
| Space group | I4/mmm | I4/mmm | I4/mmm | I4/mmm |
| a (Å) | 3.917 | 4.169 | 4.121 | 4.131 |
| c (Å) | 13.297 | 13.473 | 13.575 | 13.481 |
| Physical properties | Superconductor ($T_c$ = 38K) | Antiferromagnet ($T_N$ = 625K) | Semiconductor | Ferromagnet ($T_C$ = 180K) |

superconductor interface [20]. In conclusion, we presented the synthesis of a new ferromagneticDMS system with $T_C$ up to 180 K developed over the entire volume. Availability of bulk
specimens, independent spin and charge controls, and perfect lattice matching with the 122 FeAs superconductors and relevant antiferromagnets make this promising system decisively different from existing DMS systems based on III-V semiconductors.

Materials and Methods:

Polycrystal specimens of (Ba,K)(Zn,Mn)$_2$As$_2$ were synthesized using the arc melting solid-state reaction method similar to that described in [7]. The starting materials, namely, precursors of BaAs, KAs, ZnAs, and high-purity Mn powders, were mixed according to the nominal composition of (Ba,K)(Zn,Mn)$_2$As$_2$. The mixture was sealed inside an evacuated tantalum tube that is, in turn, sealed inside an evacuated quartz tube. The mixture was heated until 750 °C at 3 °C/min. Then, the temperature was maintained for 20 h before it was slowly decreased to room temperature at a rate of 2 °C/min.
Samples were characterized by X-ray powder diffraction with a Philips X'pert diffractometer using Cu K-edge radiation. The DC magnetic susceptibility was characterized using a superconducting quantum interference device magnetometer (Quantum Design, Inc.), whereas the electronic-transport were measured using a physical property measuring system. Neutron powder diffraction measurements were performed at the FRM-II reactor in TU Munich using the SPODI spectrometer (for details of SPODI,see http://www.sciencedirect.com/science/article/pii/S0168900211021383 .)

# Acknowledgement

The present work was supported by the Chinese NSF and Ministry of Science andTechnology (MOST) through research projects at IOP; the National Basic Research Program of China (973 Program) under grant no.2011CBA00103 at IOP and Zhejiang; the US NSF PIRE (Partnership for International Research and Education: OISE-0968226) and DMR- 1105961 projects at Columbia; the JAEA Reimei project at IOP, Columbia, PSI, McMaster and TU Munich; and NSERC and CIFAR at McMaster. We would like to thank I. Mirebeau, S.Maekawa and L. Yu for helpful discussions.

## Author contributions

Proposal of the material: C.Q.J.; research coordination: C.Q.J. and Y.J.U.; synthesis, transport and magnetization: C.Q.J., K.Z., Z.D., X.C.W., W.H., J.L.Z., X.L., Q.Q.L. and R.C.Y.; X-ray diffraction: H.D., G.M.L. and K.Z.; neutron scattering: A.S., S.R.D., P.B., Y.J.U. and B.F.; MuSR measurements: T.G., B.F., L.L., F.N., Y.J.U., H.L., E.M.; manuscript: K.Z., C.Q.J. and Y.J.U.

Competing financial interests: The authors declare no competing financial interests.

## Supplementary Information:

Stable structure of (Ba,K)(Zn,Mn)$_2$As$_2$ system:

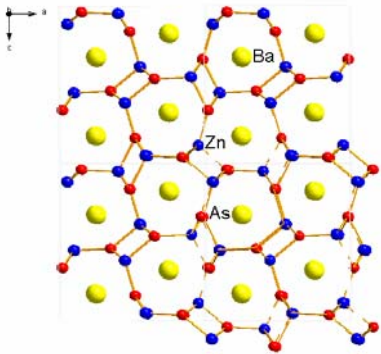

*Supplementary Figure S1. (a) Crystal structure of BaZn$_2$As$_2$ belonging to orthorhombic structure with space group Pnma. The parent compound BaZn$_2$As$_2$ crystallizes into a orthorhombic structure with space group Pnma as the room temperature phase, and the ThCr$_2$Si$_2$ structure is the stable structure of BaZn$_2$As$_2$ synthetized at high temperature (>900$^o$C).*

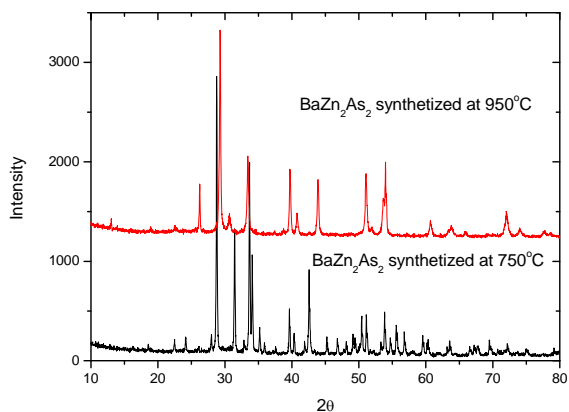

*Supplementary Figure S1.(b) X-ray diffraction (XRD) pattern of BaZn$_2$As$_2$ synthetized at 750$^o$C and 950$^o$C, respectively.*

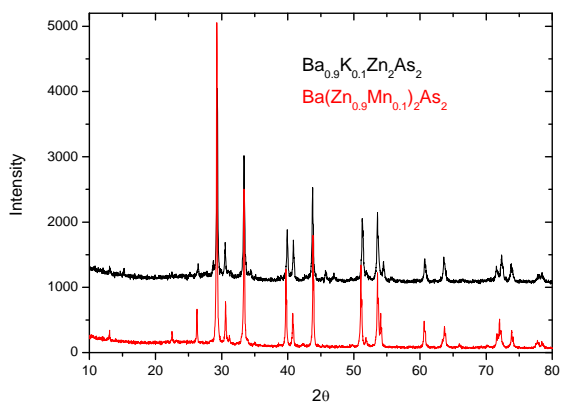

*Supplementary Figure S2. X-ray diffraction (XRD) pattern of $Ba_{0.9}K_{0.1}Zn_2As_2$ and $Ba(Zn_{0.9}Mn_{0.1})_2As_2$ synthetized at $750^oC$. Either 10% of K or Mn doping dramatically stablize the tetragonal $ThCr_2Si_2$ structure at room temperature.*